\documentclass[prl,twocolumn,showpacs,preprintnumbers]{revtex4-1}
\usepackage{graphicx}
\usepackage{amsmath}
\usepackage{amssymb}
\usepackage{times}
\usepackage{color}
\usepackage{ulem}

\def\d{\delta}

\def\S{\Sigma}
\def\s{\sigma}

\def\w{\omega}
\def\ua{\uparrow}
\def\da{\downarrow}

\def\Vec#1{\mathbf #1}

\newcommand{\OP}{\text{OP}_{i}}
\newcommand{\OPij}{\text{OP}_{ij}}
\newcommand{\OPkk}{\text{OP}_{\Vec{k}\Vec{k}'}}
\newcommand{\nb}{\bar{n}}
\newcommand{\OPb}{\overline{\text{OP}}}
\begin{document}

\title{Superconductivity on a Quasiperiodic Lattice: Extended-to-Localized Crossover of Cooper Pairs}

\author{Shiro Sakai$^1$, Nayuta Takemori$^1$, Akihisa Koga$^2$, and Ryotaro Arita$^1$}

\affiliation{
$^1$Center for Emergent Matter Science, RIKEN, Wako, Saitama 351-0198, Japan\\
$^2$Department of Physics, Tokyo Institute of Technology, Meguro, Tokyo 152-8551, Japan
}
\date{\today}
\begin{abstract}
We study a possible superconductivity in quasiperiodic systems, by portraying the issue within the attractive Hubbard model on a Penrose lattice.
Applying a real-space dynamical mean-field theory to the model consisting of 4181 sites, 
we find a superconducting phase at low temperatures.
Reflecting the nonperiodicity of the Penrose lattice, the superconducting state exhibits an  inhomogeneity. According to the type of the inhomogeneity, the superconducting phase is categorized into three different regions which cross over each other.
Among them, the weak-coupling region exhibits spatially extended Cooper pairs, which are nevertheless distinct from the conventional pairing of two electrons with opposite momenta. 
\end{abstract}
\maketitle

Quasicrystal is a crystal without translational symmetry. 
Prominent spots observed in its diffraction pattern manifest an orderly structure while they do not conform to any periodicity.
An example of such structures holds the icosahedral point-group symmetry, as first discovered by Shechtman {\it et al.} \cite{shechtman84}, and various other structures have hitherto been reported \cite{tsai87,tsai00,steurer08}.
These structures may originate novel electronic properties distinct from those of conventional periodic crystals.
In fact, previous theoretical works revealed various nontrivial properties, such as the presence of a confined state \cite{kohmoto86PRL,arai88}, fractal dimensions \cite{sutherland86,tsunetsugu86,tokihiro88}, singular continuous spectral measure \cite{kohmoto83,ostlund83,tsunetsugu86}, pseudogap in the density of states \cite{fujiwara89}, and a conductance decaying in power of system size \cite{tsunetsugu91PRB1,tsunetsugu91PRB2},
 for free electrons on the quasiperiodic lattices. 
Moreover, recent observation of quantum critical behavior in Au$_{51}$Al$_{34}$Yb$_{15}$ \cite{deguchi12} has stimulated theoretical studies \cite{watanabe13,shaginyan13,takemori15,takemura15,watanabe16,otsuki16,shinzaki16} on the role of electron correlations in these systems.

Another interesting recent observation is a superconductivity in approximant crystals (i.e.,  periodic crystals with the same local structure as the quasicrystals), Au$_{64}$Ge$_{22}$Yb$_{14}$ and Au$_{63.5}$Ge$_{20.5}$Yb$_{16}$ \cite{deguchi15}.
A superconductivity has also been reported in Al-Cu-(Mg, Li) quasicrystalline alloys \cite{wong87,wagner88}.
These observations raise fundamental questions about a possible superconductivity in quasicrystals: How can a superconductivity emerge in a system without translational symmetry? 
If it exists, what differs from the superconductivity in periodic systems?
These questions also have a relevance to experiment of ultracold atomic gases, for which 
 optical quasiperiodic lattices have been available \cite{guidoni97,guidoni99,sanchez05}.
 
According to an early consideration by Anderson \cite{anderson59} about the impurity effect on superconductivity, Cooper pairs can exist in principle even in the absence of the translational symmetry.
In this case, an electron finds its partner in the time-reversed state, which is a generalization of the standard pairing of $\Vec{k}\ua$ and $-\Vec{k}\da$.
However, as a matter of course, this does not guarantee the presence of superconductivity in quasiperiodic systems. This many-body problem requires an explicit calculation taking into account both the pairing interaction and the lattice geometry.

In this Letter, we address the above issues in a simple setting, i.e., the attractive Hubbard model on a Penrose lattice \cite{penrose74}.
On periodic lattices, the attractive Hubbard model is known to show the superconductivity at any finite value of the attraction $U<0$ in the ground state while the character of the superconducting transition changes with $U$ \cite{micnas90}:
For small values of $|U|$ (typically smaller than the bandwidth), it follows well the Bardeen-Cooper-Schrieffer (BCS) theory \cite{bardeen57} while for larger values of $|U|$ it behaves like a Bose-Einstein condensation (BEC) of incoherent pairs preformed above the transition temperature.

\begin{figure}[tb]
\center{
\includegraphics[width=0.48\textwidth]{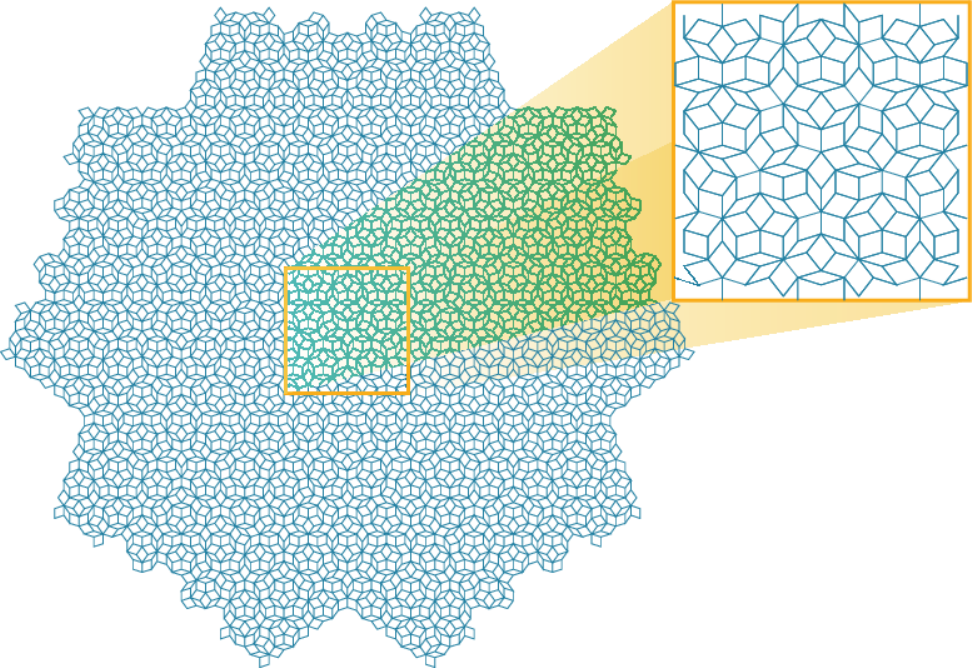}}
\caption{Two-dimensional Penrose lattice of 4181 sites. The sites are located at vertices of rhombuses.
Top-right panel is an enlarged view of a part of the lattice.
}
\label{fig:penrose}
\end{figure}

We study the attractive Hubbard model on the Penrose lattice by means of the real-space dynamical mean-field theory (RDMFT) \cite{metzner89,georges96,potthoff99,takemori15} explained below.
At low temperatures, we find a superconducting phase in a wide range of $U$ and electron density. The onsite pair amplitude shows a site dependence, constituting an inhomogeneous superconducting state. Its spatial pattern shows various orderly structures depending on the electron density and $U$. Analyzing these structures, we find three distinct regions, which cross over each other, within the superconducting phase; 
i) weak-coupling region, where the Cooper pairs are spatially extended,
ii) low-density and strong-coupling region, where the BEC picture holds aside from the electron-density modulation, and 
iii) high-density and strong-coupling region, where the density modulation is relatively weak and the Cooper pairs are short-ranged.
We show that the superconducting state i) reflects the geometry of the Penrose lattice most strongly and constitutes a new pairing state which departs from both the BCS and BEC pictures.

The Hubbard Hamiltonian reads
\begin{align}
H= t\sum_{\langle ij\rangle\s}c_{i\s}^\dagger c_{j\s}^{\phantom {\dagger}}- \mu\sum_{i\s}n_{i\s}
+U\sum_{i}n_{i\ua}n_{i\da}, 
\label{hubbard}
\end{align}
where $c_{i\s}$ ($c_{i\s}^\dagger$) annihilates (creates) an electron of spin $\s$ at site $i$ on the two-dimensional Penrose lattice and $n_{i\s}\equiv c_{i\s}^\dagger c_{i\s}$.
$t$ is the transfer integral between the nearest-neighbor sites $\langle ij\rangle$ and $\mu$ is the chemical potential.
We consider an open-boundary cluster of $N=4181$ sites, generated by iteratively applying the inflation-deflation rule \cite{levine84}.
The cluster holds a five-fold rotational symmetry as illustrated in Fig.~\ref{fig:penrose}.
We take $t=1$ as the unit of energy.
At $U=0$, the difference between the highest and lowest eigenenergies (i.e., the ``bandwidth'') is about 8.46, which is close to the value (8.47) estimated for the infinite lattice \cite{kohmoto86PRL}. 

In the RDMFT, the action of the effective impurity model is defined and solved at each symmetrically-independent site (which amounts to 444 sites for the present cluster).
For the impurity solver, we use a finite-temperature exact-diagonalization (ED) method \cite{capone07,liebsch12} extended to the superconducting state \cite{sakai15,sakai16}.
The calculated local self-energy $\S_i$ has a dependence on the site index $i$ while the nonlocal self-energies are neglected.
While it is certainly interesting to study the nonlocal correlations on the Penrose lattice, it requires a considerable methodological development: The RDMFT is one of the best currently-available methods to deal with the present model as large as 4181 sites, as well as both the BCS and BEC regimes on equal footing \cite{garg05,toschi05,bauer09,koga11,peters15,sakai15}. 
The RDMFT reproduces both the weak- and strong-coupling limits, connecting them for intermediate couplings, and thus provides a starting point for more refined theories in future.

In the superconducting state,  $\S_i$ is a 2$\times$2 Nambu matrix while here we use a plain notation just for the sake of brevity.
We define the lattice Green's function $\hat{G}_\text{lat}$ as a real-space matrix:
\begin{align}
\left[\hat{G}_\text{lat}(i\w_n)^{-1}\right]_{ij}=
\left[i\w_n \s_0+\mu\s_3-\S_i(i\w_n)\right]\d_{ij}-t\s_3\d_{\langle ij\rangle},
\label{glat}
\end{align}
where $\w_n=(2n+1)\pi T$ is the Matsubara frequency at a temperature $T$, and $\s_{0,3}$ is the Pauli matrix.
Taking the matrix inverse of the right-hand side of Eq.~(\ref{glat}), we obtain Green's function reflecting the hopping structure of the Penrose lattice. 
The dynamical mean field $g_i^0(i\w_n)$ at each site $i$ is determined self-consistently by using the diagonal component of $\hat{G}_\text{lat}(i\w_n)$ as
\begin{align}
g_i^0(i\w_n)=\left\{ \left[\hat{G}_\text{lat}(i\w_n)^{-1}\right]_{ii} + \S_i(i\w_n)\right\}^{-1}.
\label{g0}
\end{align}
In order to apply ED, we fit $g_i^0(i\w_n)$ with a function involving six bath sites, which indeed give a sufficiently accurate fitting for the parameters studied in this Letter.
We denote by $n_i$ and $\OP$ the expectation value of the electron density $\langle \sum_\s c_{i\s}^\dagger c_{i\s}\rangle$ and the onsite superconducting order parameter $\langle c_{i\ua}c_{i\da}\rangle$, respectively.
We define the site-averaged quantity $\overline{Q}\equiv\frac{1}{N}\sum_{i=1,\dots,N}Q_i$ for $Q=n$ and OP.

\begin{figure}[tb]
\center{
\includegraphics[width=0.48\textwidth]{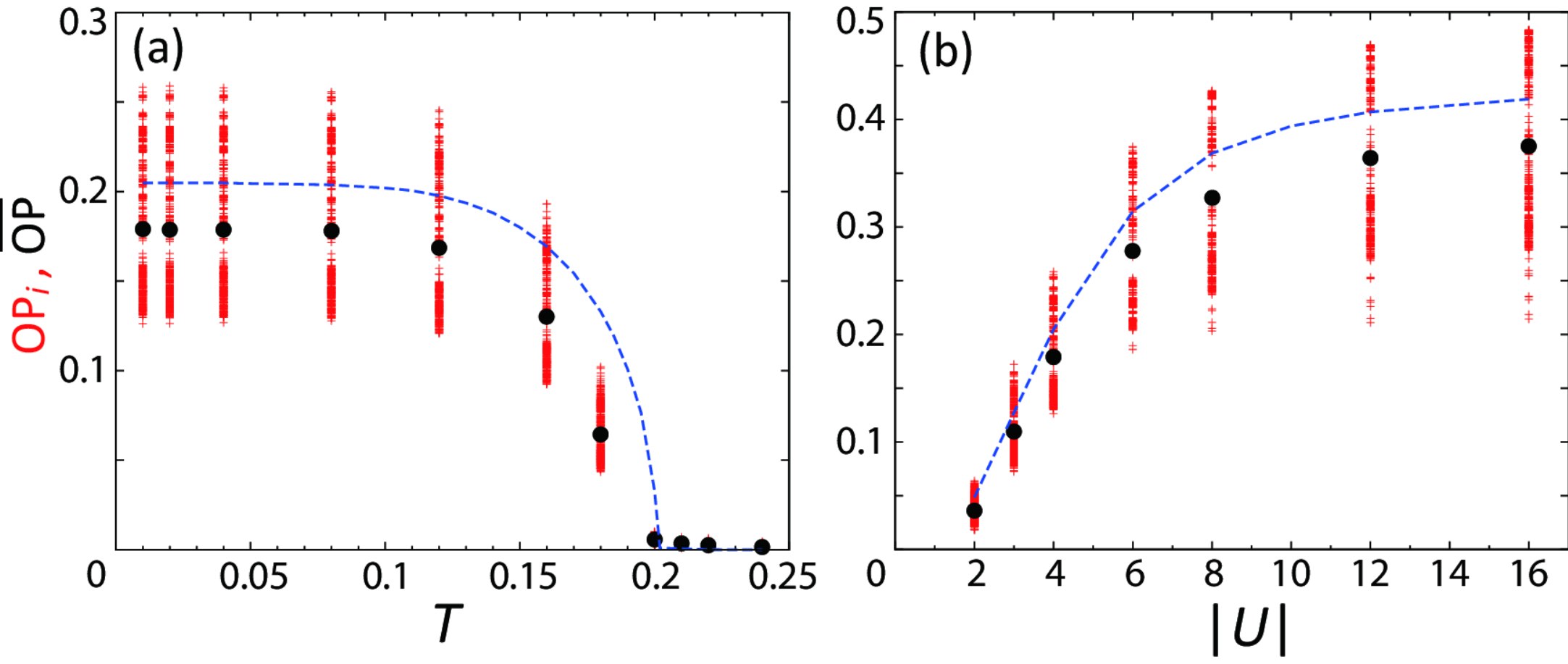}}
\caption{(a) $T$ dependence of $\OP$ (red crosses) and $\OPb$ (black dots) for $\nb=0.5$ and $U=-4$. Blue curve plots the DMFT results for the Bethe lattice with the bandwidth $8t$ at quarter filling.
(b) The same for $U$ dependence for $\nb=0.5$ and $T=0.01$.
}
\label{fig:OP-TU}
\end{figure}

Figure \ref{fig:OP-TU} shows that $\OP$ indeed becomes finite at low $T$.
The phase of $\OP$ is always found to be uniform in space (so that we take all $\OP$'s to be positive hereafter) while its intensity depends on sites as represented by the scattering red crosses.
Despite the inhomogeneity in $\OP$, the normal-to-superconducting transition occurs at the same temperature for every site within the precision of the present calculation [panel (a)].
The mean value $\OPb$ (denoted by black dots) increases monotonically with $|U|$, similarly to the case of the Bethe lattice in infinite dimensions (blue dashed curve) [panel (b)].
Note that the transition temperature and $\OPb$ show scales similar to those of the Bethe lattice with a similar bandwidth.

\begin{figure}[tb]
\center{
\includegraphics[width=0.48\textwidth]{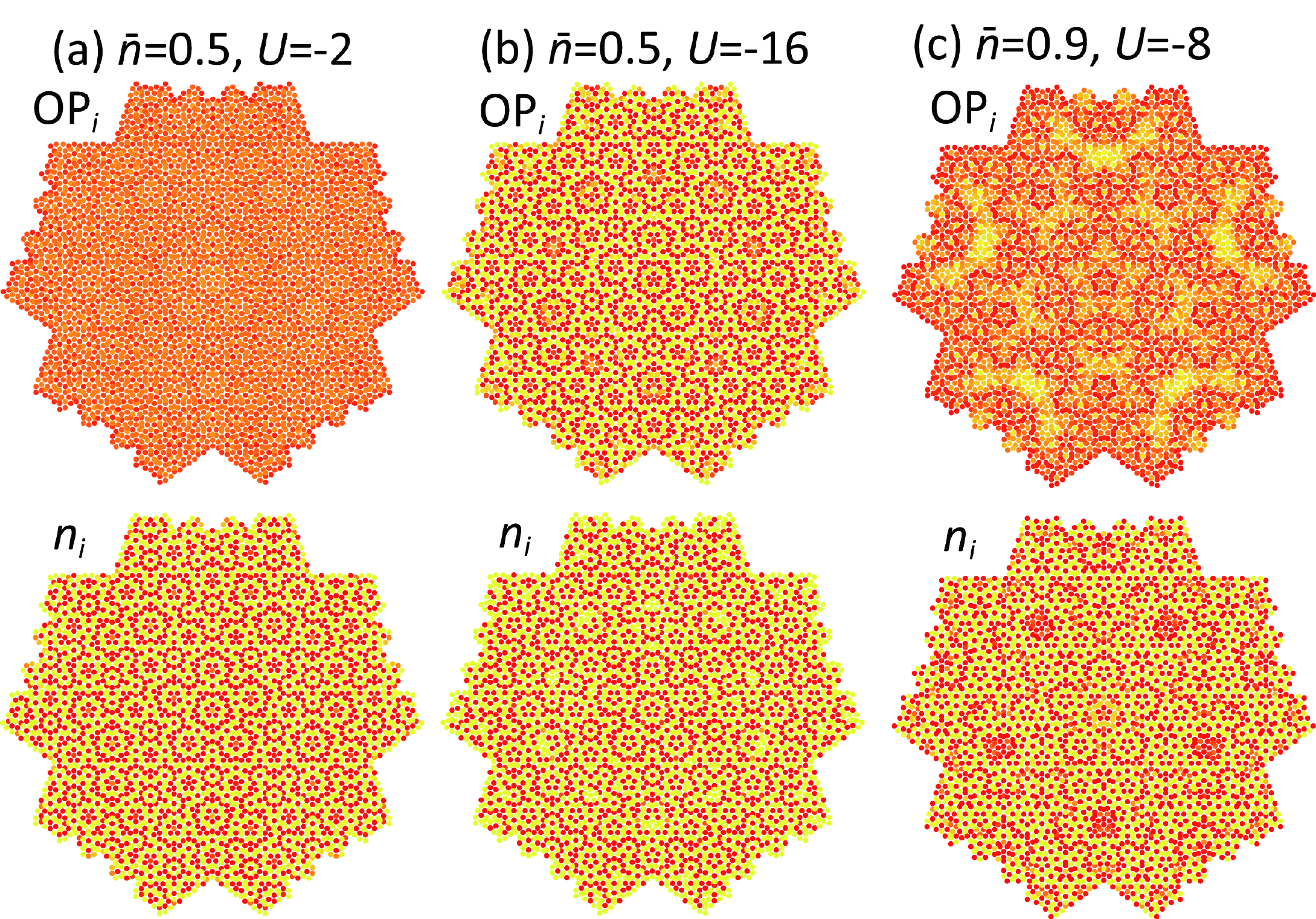}}
\caption{Spatial patterns of $\OP$ and $n_i$ at $T=0.01$ for three different sets of $U$ and $\nb$. 
The sites with $Q_i >\overline{Q}$ ($Q_i <\overline{Q}$) with $Q=\text{OP}$ and $n$ are 
colored by red (yellow).}
\label{fig:OP-r}
\end{figure}

Figure \ref{fig:OP-r} depicts the spatial patterns of $\OP$, as well as of $n_i$, at $T=0.01$.
The patterns change with $\nb$ and $U$, holding a five-fold rotational symmetry.
Here we show three distinct examples.
Panel (a) shows the result for a weak coupling, where $\OPb$ is not large [see Fig.~\ref{fig:OP-TU}(b)]. While $n_i$ clearly shows an orderly structure, $\OP$ does not show an appreciable pattern.
Panel (b) shows the result for a relatively low density and strong coupling, where $\OP$ oscillates in a short length (of the order of the distance between neighboring sites) and its spatial pattern well corresponds to that of $n_i$.
Panel (c) represents the result for a relatively high density and strong coupling, where $\OP$ oscillates in a longer length (of the order of ten sites) and shows a pattern different from that of $n_i$ though some correlation between them is still recognizable.
This result indicates a presence of another factor, in addition to $n_i$, determinant to the spatial structure of $\OP$.

\begin{figure}[tb]
\center{
\includegraphics[width=0.48\textwidth]{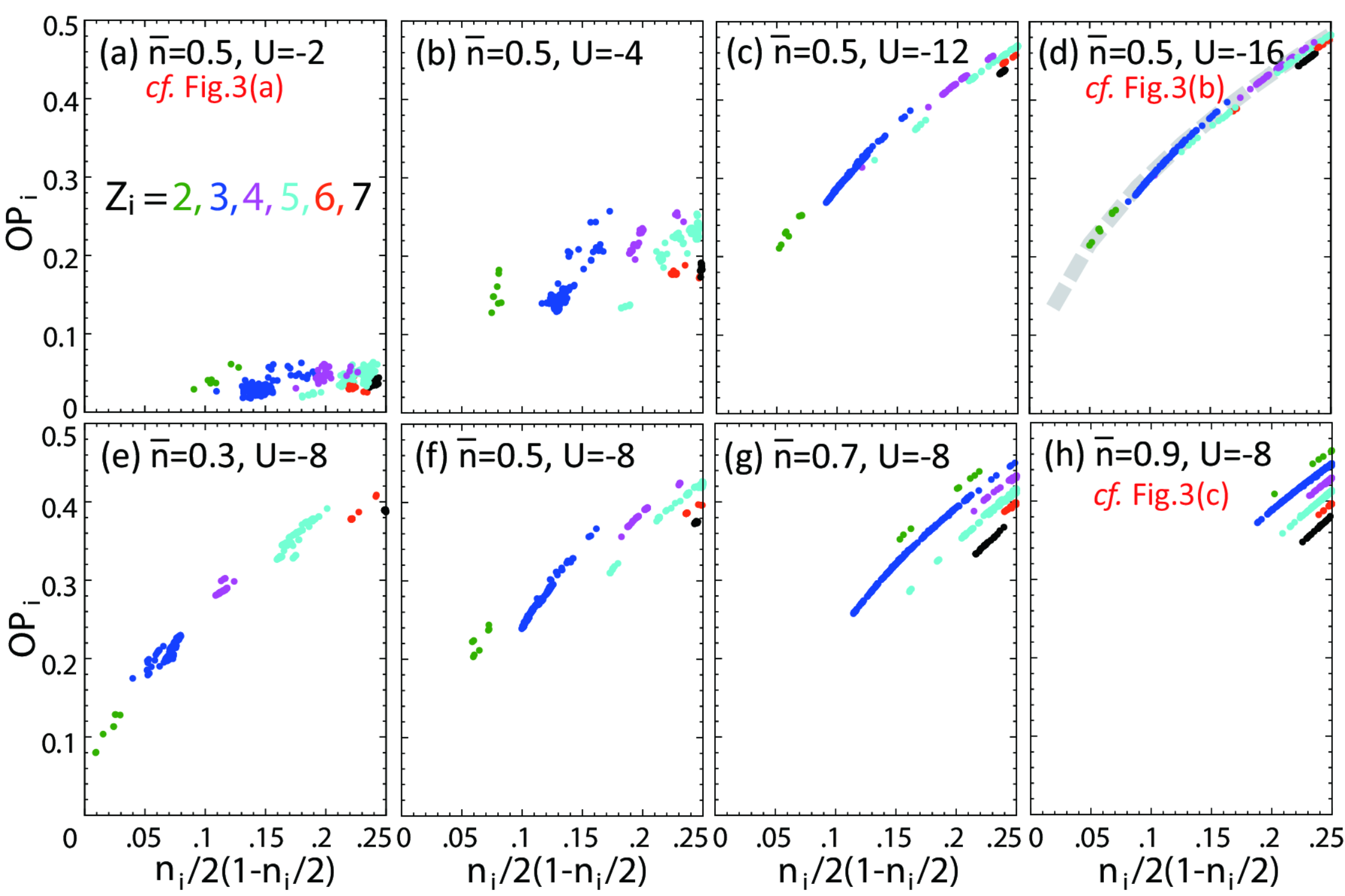}}
\caption{$\OP$ plotted against $n_i/2 (1-n_i/2)$ for various sets of $\nb$ and $U$ at $T=0.01$.
In order of (a)-(b)-(f)-(c)-(d), $|U|$ increases at a fixed $\nb=0.5$.
In order of (e)-(f)-(g)-(h), $\nb$ increases at a fixed $U=-8$.
Thick gray dashed curve in panel (d) shows the result calculated for the infinite-dimensional Bethe lattice with the bandwidth $8t$.
}
\label{fig:op-ph}
\end{figure}

This second factor is the number of bonds (the coordination number) $Z_i$ at each site, which varies in a range from two to seven, depending on the local geometry of each site.
To see this, in Fig.~\ref{fig:op-ph}, we first plot $\OP$ against the product of the particle and hole densities, $n_i/2 (1-n_i/2)$, which is apparently related to 
the density of the Bogoliubov quasiparticles \cite{ghosal98,ghosal01};
This quantity takes the maximum (0.25) for $n_i=1$ (half filling) and decreases monotonously as $n_i$ goes away from 1.
We then categorize the data with respect to $Z_i$.
For all values of $\nb$ and $U$, the data points look well grouped by $Z_i$ while they exhibit several different characteristics depending on $\nb$ and $U$.

For small $|U|$ [panel (a)], the data points are rather scattered while $n_i/2 (1-n_i/2)$ tends to be larger for larger $Z_i$.
$\OP$ does not much depend on $n_i$ and $Z_i$, in consistency with the absence of an orderly pattern in Fig.~\ref{fig:OP-r}(a).

As $|U|$ increases in order of panels (a)-(b)-(f)-(c)-(d), all the data points merge into one curve, which increases monotonously with $n_i/2 (1-n_i/2)$.
While the extent of $n_i/2 (1-n_i/2)$ remains to be similar ($0.05\lesssim n_i < 0.25$) for all the studied values of $U$, $\OP$ ranges considerably wider for larger $|U|$; 
$0.12\lesssim \OP \lesssim 0.26$ for $U=-4$ and $0.2\lesssim \OP \lesssim 0.5$ for $U=-16$. 

At $U=-16$, each section of the curve is well grouped by $Z_i$: 
There remains a tendency that $n_i/2 (1-n_i/2)$ is larger for larger $Z_i$. 
This suggests that $n_i$ is determined mainly by $Z_i$ while $\OP$ is governed by $n_i/2 (1-n_i/2)$ (or more simply $n_i$) rather than $Z_i$ itself.
This is likely because the Cooper pairs at $U=-16$ are so strongly localized that $Z_i$ is not directly relevant.
In panel (d), we also plot the result for the infinite-dimensional Bethe lattice (with bandwidth $8t$) at $U=-16$ and $T=0.01$, where the electron density is changed from 0.04 to 1.0 (gray dashed curve). 
Its nice agreement with the Penrose results confirms that $\OP$ in this region follows the behavior expected from a local physics controlled only by the electron density.
Namely, provided an inhomogeneity in the electron density,  the superconductivity in this region is well understood within the BEC picture. 

In turn, this demonstrates that the superconductivity at {\it smaller} $|U|$ [panels (a) and (b)],
where the data points do not follow a simple curve, is not determined solely by the local physics and instead reflects the geometry around each site. 
Namely, the Cooper pairs in this region are extended in space. 
Since the lack of the translational symmetry in the Penrose lattice does not allow the conventional Cooper pairing formed at the opposite Fermi momenta, the extended Cooper pairs in this weak-coupling regime [depicted in Fig.~\ref{fig:OP-r}(a)] should be unconventional. We shall substantiate this point below.

By changing $\nb$ in order of Figs.~\ref{fig:OP-r}(e)-(f)-(g)-(h), we find even different characteristics in the $\OP$-$n_i/2 (1-n_i/2)$ plot.
The structure at $\nb=0.3$ [panel (e)] is similar to panel (d) discussed just above.
Then, as $\nb$ increases with fixing $U=-8$, the data points overall shift to a higher value of $n_i/2 (1-n_i/2)$, and eventually at $\nb=0.9$, most of the points are within $0.2\lesssim n_i/2 (1-n_i/2)\leq 0.25$, i.e., \{$n_i$\} approaches a homogeneous distribution.

Here a caveat is needed: In general a charge order due to the attractive interaction can occur for $\nb\simeq 1$ while we have suppressed it by mixing $g_i^0(i\w_n)$ in Eq.~(\ref{g0}) with the one obtained in the previous self-consistency loop. 
This is to focus on the effect of inhomogeneity inherent to quasiperiodicity rather than highlighting the effect of the charge ordering which may occur particularly easily in the present bipartite lattice.
In the limit of strong coupling, this result would be connected to the antiferromagnetic phase found in the Heisenberg model on the Penrose lattice \cite{jagannathan07}.

In Fig.~\ref{fig:op-ph}(h), a comparison at a fixed $n_i$ shows that $\OP$ decreases with $Z_i$. 
This decrease can be explained by the curve in Fig.~\ref{fig:OP-TU}(b): The sites with large $Z_i$ are considered to be weakly correlated compared to those with smaller $Z_i$, and then, according to Fig.~\ref{fig:OP-TU}(b), the former $\OP$ is smaller than the latter.
Thus, $\OP$ in Fig.~\ref{fig:op-ph}(h) is determined by $Z_i$ rather than $n_i$ and this is again beyond the local physics. 
On the other hand, the fact that for each $Z_i$ the points follow one curve indicates the short-ranged pairs:
The geometry beyond the nearest neighbors does not play a significant role (other than changing $n_i$ slightly).   
This is distinct from Fig.~\ref{fig:op-ph}(a), where a longer-range geometry beyond the nearest neighbors plays a role. 
We therefore conclude that the superconducting state depicted in Figs.~\ref{fig:op-ph}(h) and \ref{fig:OP-r}(c) constitutes another type of unconventional superconductivity.

\begin{figure}[tb]
\center{
\includegraphics[width=0.48\textwidth]{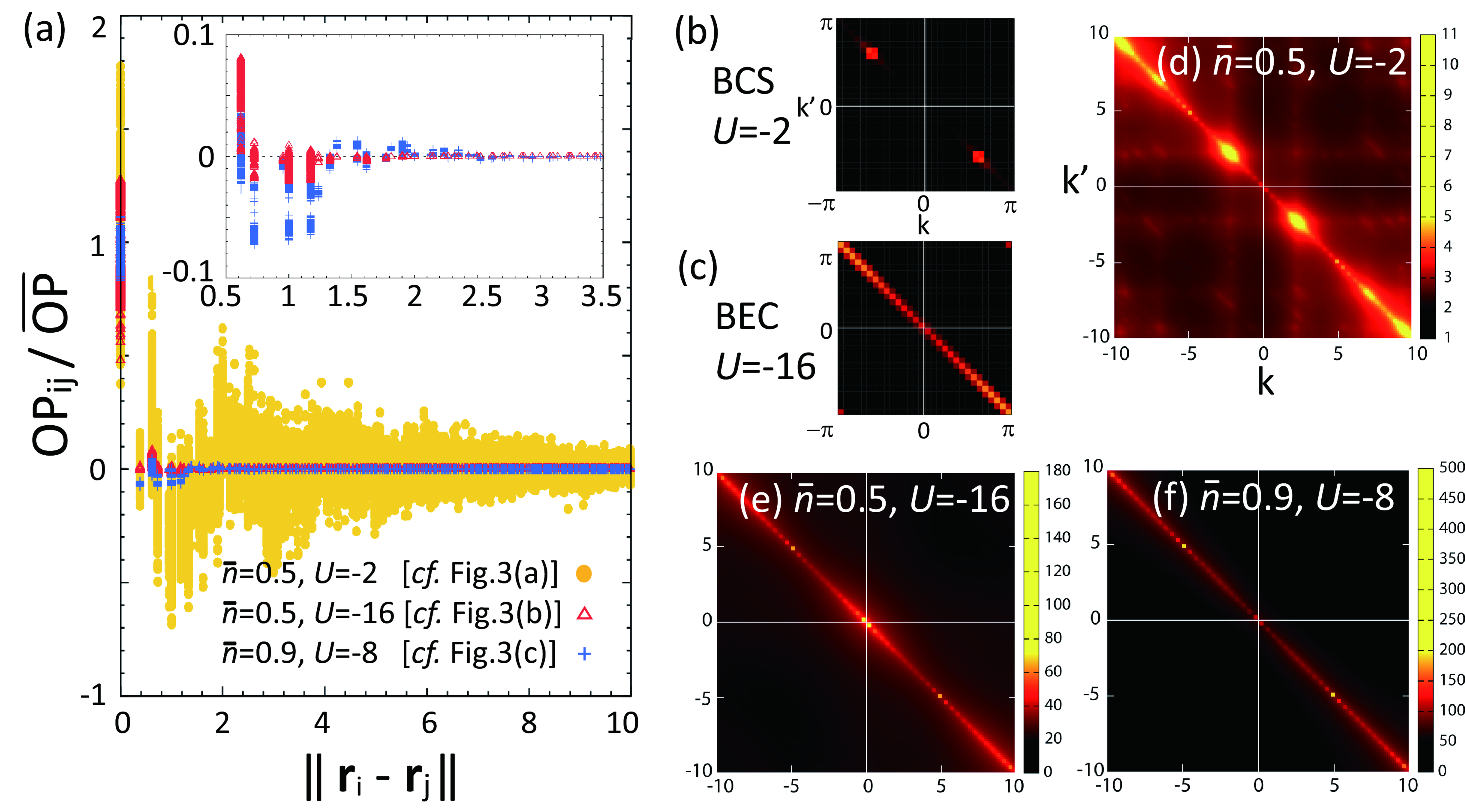}}
\caption{(a) Offsite pair amplitude $\OPij \equiv \langle c_{i\ua}c_{j\da}\rangle$ (normalized by $\OPb$) plotted against the Euclidean distance $|| \Vec{r}_i-\Vec{r}_j ||$ for the three states shown in Fig.~\ref{fig:OP-r}. Inset: Enlarged view for the short-distance part of red triangles and blue crosses. 
(b), (c)  Intensity map of $|\OPkk|$ calculated for a square lattice with the bandwidth $8t$ at quater filling for $U=-2$ and $-16$, respectively. Just for plotting purpose, we set $k_x=k_y=k$ and $k_x'=k_y'=k'$.
(d), (e), (f) The same quantity for the three states in panel (a), plotted with a cutoff at $|k|, |k'|=10$.
}
\label{fig:F-r}
\end{figure}

In order to examine the change of the spatial extent of the Cooper pairs, we calculate the offsite
pair amplitude, $\OPij\equiv \langle c_{i\ua}c_{j\da}\rangle$.
Figure \ref{fig:F-r}(a) plots it (after the normalization by $\OPb$) against the Euclidean distance $|| \Vec{r}_i-\Vec{r}_j ||$ ($\Vec{r}_i$: the Cartesian coordinate of site $i$) for the three states in Fig.~\ref{fig:OP-r}.
For $\nb=0.5$ and $U=-2$ (yellow), $\OPij$ decays slowly, demonstrating that the Cooper pairs are extended in space.
On the other hand, for $\nb=0.5$ and $U=-16$ (red) and for $\nb=0.9$ and $U=-8$ (blue), $\OPij$ decays much faster: The inset shows that the former decays even faster than the latter.
These results support the above interpretations of the three different superconducting states.

Figures \ref{fig:F-r}(b)-(f) further clarify the nature of the Cooper pairs. Here we have defined a Fourier-transformed pair amplitude 
$\OPkk \equiv \langle c_{\Vec{k}\ua}c_{\Vec{k}'\da}\rangle$.
In periodic systems, $\OPkk$ is finite only along the $\Vec{k}'=-\Vec{k}$ line, as demonstrated in panels (b) and (c) for a square lattice:
Panel (b) shows a prototype in the BCS region where $\text{OP}_{\Vec{k},-\Vec{k}}$ is substantial only around the Fermi momenta, while panel (c) shows a prototype in the BEC region where $\OPkk$ is distributed along the $\Vec{k}'=-\Vec{k}$ line.

This latter BEC feature is found in Figs.~\ref{fig:F-r}(e) and (f) of Penrose, although $\OPkk$ is not strictly zero even for $\Vec{k}'\neq -\Vec{k}$.
Meanwhile, in Fig.~\ref{fig:F-r}(d) several high-intensity lines are discernible besides $\Vec{k}'=-\Vec{k}$.
In addition, high-intensity spots exist on the $\Vec{k}'=-\Vec{k}$ line, similarity to the BCS case [panel (b)], despite that the Fermi momentum is undefined on the Penrose lattice. 
Because a similar structure is obtained even when we use only inner sites for the Fourier transformation, this cannot be attributed to a boundary effect.
This nontrivial structure differs from a disordered BCS state, suggesting a unique pairing intrinsic to the Penrose lattice. 
Note that the lack of the inversion symmetry in the Penrose lattice suggests that this pairing is a mixture of spin singlet and triplet.

\begin{figure}[tb]
\center{
\includegraphics[width=0.45\textwidth]{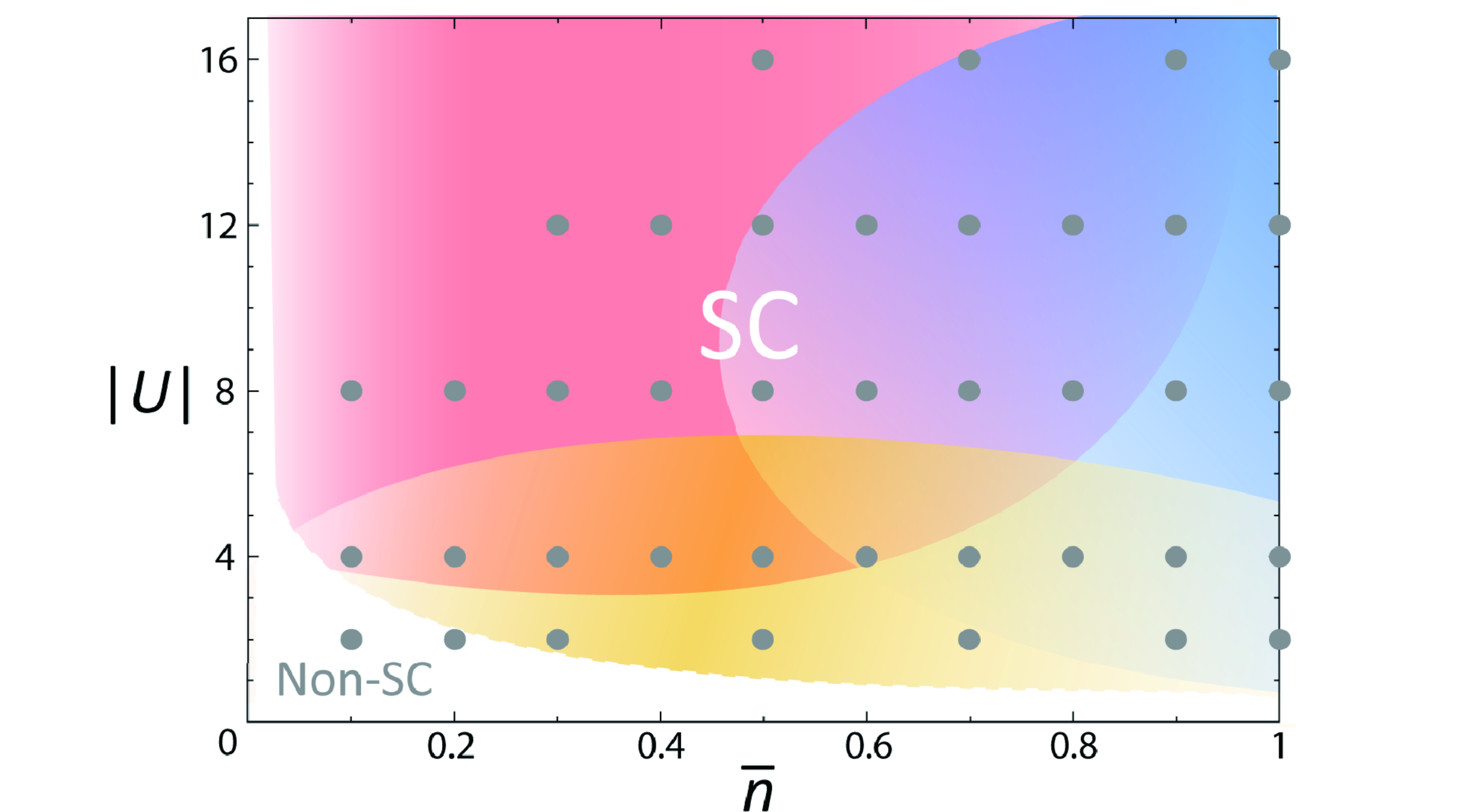}}
\caption{Phase diagram on the $\nb$-$U$ plane at $T=0.01$. SC denotes the superconducting phase. The yellow, red, and blue regions, which are judged from the characteristics seen in the $\OP$-$n_i/2 (1-n_i/2)$ plots like Fig.~\ref{fig:op-ph}, denote the superconducting states represented by Figs.~\ref{fig:OP-r}(a), (b), and (c), respectively.
}
\label{fig:pd}
\end{figure}

We summarize the results in Fig.~\ref{fig:pd}.
The phase diagram is calculated at $T=0.01$, where only the ground state has a substantial Boltzmann weight except for $\nb\lesssim0.3$ and/or $|U|\lesssim 2$.
The yellow, red, and blue regions denote the superconducting states represented by Figs.~\ref{fig:OP-r}(a), (b), and (c), respectively.
In the yellow region, the Cooper pairs are spatially extended while the pairing departs from the BCS theory. 
The red region follows the BEC picture aside from the electron-density modulation.
In the blue region, the Cooper pairs are nonlocal but short-ranged.  
These three states cross over each other, as expressed by the overlaps of colors.
 
These results reveal that a quasiperiodic system is a fertile ground of novel superconductivity, which would become particularly feracious when the fractal geometry interplays with the macroscopic quantum nature of superconductivity. Many issues remain open, such as 
a functional form of the extended non-BCS Cooper pairs, physical properties of these superconductors, and the effect of nonlocal correlations.
Studies for more realistic models, with including phonons, orbital degrees of freedom, and a variation in the transfer integrals, are also an intriguing future subject.

\begin{acknowledgments}
We thank Masahito Ueda, Masatoshi Imada and Shintaro Hoshino for stimulating discussions.
S.S. is supported by JSPS KAKENHI Grant No. JP26800179 and by MEXT KAKENHI Grant No. 16H06345.
N.T. is supported by RIKEN Special Postdoctoral Researchers Program.
A.K. is supported by MEXT KAKENHI Grant No. 25800193 and 16H01066.
R.A. is supported by MEXT KAKENHI Grant No. 15H05883 and 16H06345.
\end{acknowledgments}

\bibliography{ref}

\begin{thebibliography}{48}%
\makeatletter
\providecommand \@ifxundefined [1]{%
 \@ifx{#1\undefined}
}%
\providecommand \@ifnum [1]{%
 \ifnum #1\expandafter \@firstoftwo
 \else \expandafter \@secondoftwo
 \fi
}%
\providecommand \@ifx [1]{%
 \ifx #1\expandafter \@firstoftwo
 \else \expandafter \@secondoftwo
 \fi
}%
\providecommand \natexlab [1]{#1}%
\providecommand \enquote  [1]{``#1''}%
\providecommand \bibnamefont  [1]{#1}%
\providecommand \bibfnamefont [1]{#1}%
\providecommand \citenamefont [1]{#1}%
\providecommand \href@noop [0]{\@secondoftwo}%
\providecommand \href [0]{\begingroup \@sanitize@url \@href}%
\providecommand \@href[1]{\@@startlink{#1}\@@href}%
\providecommand \@@href[1]{\endgroup#1\@@endlink}%
\providecommand \@sanitize@url [0]{\catcode `\\12\catcode `\$12\catcode
  `\&12\catcode `\#12\catcode `\^12\catcode `\_12\catcode `\%12\relax}%
\providecommand \@@startlink[1]{}%
\providecommand \@@endlink[0]{}%
\providecommand \url  [0]{\begingroup\@sanitize@url \@url }%
\providecommand \@url [1]{\endgroup\@href {#1}{\urlprefix }}%
\providecommand \urlprefix  [0]{URL }%
\providecommand \Eprint [0]{\href }%
\providecommand \doibase [0]{http://dx.doi.org/}%
\providecommand \selectlanguage [0]{\@gobble}%
\providecommand \bibinfo  [0]{\@secondoftwo}%
\providecommand \bibfield  [0]{\@secondoftwo}%
\providecommand \translation [1]{[#1]}%
\providecommand \BibitemOpen [0]{}%
\providecommand \bibitemStop [0]{}%
\providecommand \bibitemNoStop [0]{.\EOS\space}%
\providecommand \EOS [0]{\spacefactor3000\relax}%
\providecommand \BibitemShut  [1]{\csname bibitem#1\endcsname}%
\let\auto@bib@innerbib\@empty
\bibitem [{\citenamefont {Shechtman}\ \emph {et~al.}(1984)\citenamefont
  {Shechtman}, \citenamefont {Blech}, \citenamefont {Gratias},\ and\
  \citenamefont {Cahn}}]{shechtman84}%
  \BibitemOpen
  \bibfield  {author} {\bibinfo {author} {\bibfnamefont {D.}~\bibnamefont
  {Shechtman}}, \bibinfo {author} {\bibfnamefont {I.}~\bibnamefont {Blech}},
  \bibinfo {author} {\bibfnamefont {D.}~\bibnamefont {Gratias}}, \ and\
  \bibinfo {author} {\bibfnamefont {J.~W.}\ \bibnamefont {Cahn}},\ }\href
  {\doibase 10.1103/PhysRevLett.53.1951} {\bibfield  {journal} {\bibinfo
  {journal} {Phys. Rev. Lett.}\ }\textbf {\bibinfo {volume} {53}},\ \bibinfo
  {pages} {1951} (\bibinfo {year} {1984})}\BibitemShut {NoStop}%
\bibitem [{\citenamefont {Tsai}\ \emph {et~al.}(1988)\citenamefont {Tsai},
  \citenamefont {Inoue},\ and\ \citenamefont {Masumoto}}]{tsai87}%
  \BibitemOpen
  \bibfield  {author} {\bibinfo {author} {\bibfnamefont {A.~P.}\ \bibnamefont
  {Tsai}}, \bibinfo {author} {\bibfnamefont {A.}~\bibnamefont {Inoue}}, \ and\
  \bibinfo {author} {\bibfnamefont {T.}~\bibnamefont {Masumoto}},\ }\href
  {http://stacks.iop.org/1347-4065/27/i=9A/a=L1587} {\bibfield  {journal}
  {\bibinfo  {journal} {Jpn. J. Appl. Phys.}\ }\textbf {\bibinfo {volume}
  {27}},\ \bibinfo {pages} {L1587} (\bibinfo {year} {1988})}\BibitemShut
  {NoStop}%
\bibitem [{\citenamefont {Tsai}\ \emph {et~al.}(2000)\citenamefont {Tsai},
  \citenamefont {Guo}, \citenamefont {Abe}, \citenamefont {Takakura},\ and\
  \citenamefont {Sato}}]{tsai00}%
  \BibitemOpen
  \bibfield  {author} {\bibinfo {author} {\bibfnamefont {A.~P.}\ \bibnamefont
  {Tsai}}, \bibinfo {author} {\bibfnamefont {J.~Q.}\ \bibnamefont {Guo}},
  \bibinfo {author} {\bibfnamefont {E.}~\bibnamefont {Abe}}, \bibinfo {author}
  {\bibfnamefont {H.}~\bibnamefont {Takakura}}, \ and\ \bibinfo {author}
  {\bibfnamefont {T.~J.}\ \bibnamefont {Sato}},\ }\href {\doibase
  10.1038/35046202} {\bibfield  {journal} {\bibinfo  {journal} {Nature}\
  }\textbf {\bibinfo {volume} {408}},\ \bibinfo {pages} {537} (\bibinfo {year}
  {2000})}\BibitemShut {NoStop}%
\bibitem [{\citenamefont {Steurer}\ and\ \citenamefont
  {Deloudi}(2008)}]{steurer08}%
  \BibitemOpen
  \bibfield  {author} {\bibinfo {author} {\bibfnamefont {W.}~\bibnamefont
  {Steurer}}\ and\ \bibinfo {author} {\bibfnamefont {S.}~\bibnamefont
  {Deloudi}},\ }\href {\doibase 10.1107/S0108767307038627} {\bibfield
  {journal} {\bibinfo  {journal} {Acta Crystallographica Section A}\ }\textbf
  {\bibinfo {volume} {64}},\ \bibinfo {pages} {1} (\bibinfo {year}
  {2008})}\BibitemShut {NoStop}%
\bibitem [{\citenamefont {Kohmoto}\ and\ \citenamefont
  {Sutherland}(1986)}]{kohmoto86PRL}%
  \BibitemOpen
  \bibfield  {author} {\bibinfo {author} {\bibfnamefont {M.}~\bibnamefont
  {Kohmoto}}\ and\ \bibinfo {author} {\bibfnamefont {B.}~\bibnamefont
  {Sutherland}},\ }\href {\doibase 10.1103/PhysRevLett.56.2740} {\bibfield
  {journal} {\bibinfo  {journal} {Phys. Rev. Lett.}\ }\textbf {\bibinfo
  {volume} {56}},\ \bibinfo {pages} {2740} (\bibinfo {year}
  {1986})}\BibitemShut {NoStop}%
\bibitem [{\citenamefont {Arai}\ \emph {et~al.}(1988)\citenamefont {Arai},
  \citenamefont {Tokihiro}, \citenamefont {Fujiwara},\ and\ \citenamefont
  {Kohmoto}}]{arai88}%
  \BibitemOpen
  \bibfield  {author} {\bibinfo {author} {\bibfnamefont {M.}~\bibnamefont
  {Arai}}, \bibinfo {author} {\bibfnamefont {T.}~\bibnamefont {Tokihiro}},
  \bibinfo {author} {\bibfnamefont {T.}~\bibnamefont {Fujiwara}}, \ and\
  \bibinfo {author} {\bibfnamefont {M.}~\bibnamefont {Kohmoto}},\ }\href
  {\doibase 10.1103/PhysRevB.38.1621} {\bibfield  {journal} {\bibinfo
  {journal} {Phys. Rev. B}\ }\textbf {\bibinfo {volume} {38}},\ \bibinfo
  {pages} {1621} (\bibinfo {year} {1988})}\BibitemShut {NoStop}%
\bibitem [{\citenamefont {Sutherland}(1986)}]{sutherland86}%
  \BibitemOpen
  \bibfield  {author} {\bibinfo {author} {\bibfnamefont {B.}~\bibnamefont
  {Sutherland}},\ }\href {\doibase 10.1103/PhysRevB.34.3904} {\bibfield
  {journal} {\bibinfo  {journal} {Phys. Rev. B}\ }\textbf {\bibinfo {volume}
  {34}},\ \bibinfo {pages} {3904} (\bibinfo {year} {1986})}\BibitemShut
  {NoStop}%
\bibitem [{\citenamefont {Tsunetsugu}\ \emph {et~al.}(1986)\citenamefont
  {Tsunetsugu}, \citenamefont {Fujiwara}, \citenamefont {Ueda},\ and\
  \citenamefont {Tokihiro}}]{tsunetsugu86}%
  \BibitemOpen
  \bibfield  {author} {\bibinfo {author} {\bibfnamefont {H.}~\bibnamefont
  {Tsunetsugu}}, \bibinfo {author} {\bibfnamefont {T.}~\bibnamefont
  {Fujiwara}}, \bibinfo {author} {\bibfnamefont {K.}~\bibnamefont {Ueda}}, \
  and\ \bibinfo {author} {\bibfnamefont {T.}~\bibnamefont {Tokihiro}},\ }\href
  {\doibase 10.1143/JPSJ.55.1420} {\bibfield  {journal} {\bibinfo  {journal}
  {J. Phys. Soc. Jpn.}\ }\textbf {\bibinfo {volume} {55}},\ \bibinfo {pages}
  {1420} (\bibinfo {year} {1986})}\BibitemShut {NoStop}%
\bibitem [{\citenamefont {Tokihiro}\ \emph {et~al.}(1988)\citenamefont
  {Tokihiro}, \citenamefont {Fujiwara},\ and\ \citenamefont
  {Arai}}]{tokihiro88}%
  \BibitemOpen
  \bibfield  {author} {\bibinfo {author} {\bibfnamefont {T.}~\bibnamefont
  {Tokihiro}}, \bibinfo {author} {\bibfnamefont {T.}~\bibnamefont {Fujiwara}},
  \ and\ \bibinfo {author} {\bibfnamefont {M.}~\bibnamefont {Arai}},\ }\href
  {\doibase 10.1103/PhysRevB.38.5981} {\bibfield  {journal} {\bibinfo
  {journal} {Phys. Rev. B}\ }\textbf {\bibinfo {volume} {38}},\ \bibinfo
  {pages} {5981} (\bibinfo {year} {1988})}\BibitemShut {NoStop}%
\bibitem [{\citenamefont {Kohmoto}\ \emph {et~al.}(1983)\citenamefont
  {Kohmoto}, \citenamefont {Kadanoff},\ and\ \citenamefont {Tang}}]{kohmoto83}%
  \BibitemOpen
  \bibfield  {author} {\bibinfo {author} {\bibfnamefont {M.}~\bibnamefont
  {Kohmoto}}, \bibinfo {author} {\bibfnamefont {L.~P.}\ \bibnamefont
  {Kadanoff}}, \ and\ \bibinfo {author} {\bibfnamefont {C.}~\bibnamefont
  {Tang}},\ }\href {\doibase 10.1103/PhysRevLett.50.1870} {\bibfield  {journal}
  {\bibinfo  {journal} {Phys. Rev. Lett.}\ }\textbf {\bibinfo {volume} {50}},\
  \bibinfo {pages} {1870} (\bibinfo {year} {1983})}\BibitemShut {NoStop}%
\bibitem [{\citenamefont {Ostlund}\ \emph {et~al.}(1983)\citenamefont
  {Ostlund}, \citenamefont {Pandit}, \citenamefont {Rand}, \citenamefont
  {Schellnhuber},\ and\ \citenamefont {Siggia}}]{ostlund83}%
  \BibitemOpen
  \bibfield  {author} {\bibinfo {author} {\bibfnamefont {S.}~\bibnamefont
  {Ostlund}}, \bibinfo {author} {\bibfnamefont {R.}~\bibnamefont {Pandit}},
  \bibinfo {author} {\bibfnamefont {D.}~\bibnamefont {Rand}}, \bibinfo {author}
  {\bibfnamefont {H.~J.}\ \bibnamefont {Schellnhuber}}, \ and\ \bibinfo
  {author} {\bibfnamefont {E.~D.}\ \bibnamefont {Siggia}},\ }\href {\doibase
  10.1103/PhysRevLett.50.1873} {\bibfield  {journal} {\bibinfo  {journal}
  {Phys. Rev. Lett.}\ }\textbf {\bibinfo {volume} {50}},\ \bibinfo {pages}
  {1873} (\bibinfo {year} {1983})}\BibitemShut {NoStop}%
\bibitem [{\citenamefont {Fujiwara}(1989)}]{fujiwara89}%
  \BibitemOpen
  \bibfield  {author} {\bibinfo {author} {\bibfnamefont {T.}~\bibnamefont
  {Fujiwara}},\ }\href {\doibase 10.1103/PhysRevB.40.942} {\bibfield  {journal}
  {\bibinfo  {journal} {Phys. Rev. B}\ }\textbf {\bibinfo {volume} {40}},\
  \bibinfo {pages} {942} (\bibinfo {year} {1989})}\BibitemShut {NoStop}%
\bibitem [{\citenamefont {Tsunetsugu}\ \emph {et~al.}(1991)\citenamefont
  {Tsunetsugu}, \citenamefont {Fujiwara}, \citenamefont {Ueda},\ and\
  \citenamefont {Tokihiro}}]{tsunetsugu91PRB1}%
  \BibitemOpen
  \bibfield  {author} {\bibinfo {author} {\bibfnamefont {H.}~\bibnamefont
  {Tsunetsugu}}, \bibinfo {author} {\bibfnamefont {T.}~\bibnamefont
  {Fujiwara}}, \bibinfo {author} {\bibfnamefont {K.}~\bibnamefont {Ueda}}, \
  and\ \bibinfo {author} {\bibfnamefont {T.}~\bibnamefont {Tokihiro}},\ }\href
  {\doibase 10.1103/PhysRevB.43.8879} {\bibfield  {journal} {\bibinfo
  {journal} {Phys. Rev. B}\ }\textbf {\bibinfo {volume} {43}},\ \bibinfo
  {pages} {8879} (\bibinfo {year} {1991})}\BibitemShut {NoStop}%
\bibitem [{\citenamefont {Tsunetsugu}\ and\ \citenamefont
  {Ueda}(1991)}]{tsunetsugu91PRB2}%
  \BibitemOpen
  \bibfield  {author} {\bibinfo {author} {\bibfnamefont {H.}~\bibnamefont
  {Tsunetsugu}}\ and\ \bibinfo {author} {\bibfnamefont {K.}~\bibnamefont
  {Ueda}},\ }\href {\doibase 10.1103/PhysRevB.43.8892} {\bibfield  {journal}
  {\bibinfo  {journal} {Phys. Rev. B}\ }\textbf {\bibinfo {volume} {43}},\
  \bibinfo {pages} {8892} (\bibinfo {year} {1991})}\BibitemShut {NoStop}%
\bibitem [{\citenamefont {Deguchi}\ \emph {et~al.}(2012)\citenamefont
  {Deguchi}, \citenamefont {Matsukawa}, \citenamefont {Sato}, \citenamefont
  {Hattori}, \citenamefont {Ishida}, \citenamefont {Takakura},\ and\
  \citenamefont {Ishimasa}}]{deguchi12}%
  \BibitemOpen
  \bibfield  {author} {\bibinfo {author} {\bibfnamefont {K.}~\bibnamefont
  {Deguchi}}, \bibinfo {author} {\bibfnamefont {S.}~\bibnamefont {Matsukawa}},
  \bibinfo {author} {\bibfnamefont {N.~K.}\ \bibnamefont {Sato}}, \bibinfo
  {author} {\bibfnamefont {T.}~\bibnamefont {Hattori}}, \bibinfo {author}
  {\bibfnamefont {K.}~\bibnamefont {Ishida}}, \bibinfo {author} {\bibfnamefont
  {H.}~\bibnamefont {Takakura}}, \ and\ \bibinfo {author} {\bibfnamefont
  {T.}~\bibnamefont {Ishimasa}},\ }\href {\doibase 10.1038/nmat3432} {\bibfield
   {journal} {\bibinfo  {journal} {Nat. Mater.}\ }\textbf {\bibinfo {volume}
  {11}},\ \bibinfo {pages} {1013} (\bibinfo {year} {2012})}\BibitemShut
  {NoStop}%
\bibitem [{\citenamefont {Watanabe}\ and\ \citenamefont
  {Miyake}(2013)}]{watanabe13}%
  \BibitemOpen
  \bibfield  {author} {\bibinfo {author} {\bibfnamefont {S.}~\bibnamefont
  {Watanabe}}\ and\ \bibinfo {author} {\bibfnamefont {K.}~\bibnamefont
  {Miyake}},\ }\href {\doibase 10.7566/JPSJ.82.083704} {\bibfield  {journal}
  {\bibinfo  {journal} {J. Phys. Soc. Jpn.}\ }\textbf {\bibinfo {volume}
  {82}},\ \bibinfo {pages} {083704} (\bibinfo {year} {2013})}\BibitemShut
  {NoStop}%
\bibitem [{\citenamefont {Shaginyan}\ \emph {et~al.}(2013)\citenamefont
  {Shaginyan}, \citenamefont {Msezane}, \citenamefont {Popov}, \citenamefont
  {Japaridze},\ and\ \citenamefont {Khodel}}]{shaginyan13}%
  \BibitemOpen
  \bibfield  {author} {\bibinfo {author} {\bibfnamefont {V.~R.}\ \bibnamefont
  {Shaginyan}}, \bibinfo {author} {\bibfnamefont {A.~Z.}\ \bibnamefont
  {Msezane}}, \bibinfo {author} {\bibfnamefont {K.~G.}\ \bibnamefont {Popov}},
  \bibinfo {author} {\bibfnamefont {G.~S.}\ \bibnamefont {Japaridze}}, \ and\
  \bibinfo {author} {\bibfnamefont {V.~A.}\ \bibnamefont {Khodel}},\ }\href
  {\doibase 10.1103/PhysRevB.87.245122} {\bibfield  {journal} {\bibinfo
  {journal} {Phys. Rev. B}\ }\textbf {\bibinfo {volume} {87}},\ \bibinfo
  {pages} {245122} (\bibinfo {year} {2013})}\BibitemShut {NoStop}%
\bibitem [{\citenamefont {Takemori}\ and\ \citenamefont
  {Koga}(2015)}]{takemori15}%
  \BibitemOpen
  \bibfield  {author} {\bibinfo {author} {\bibfnamefont {N.}~\bibnamefont
  {Takemori}}\ and\ \bibinfo {author} {\bibfnamefont {A.}~\bibnamefont
  {Koga}},\ }\href {\doibase 10.7566/JPSJ.84.023701} {\bibfield  {journal}
  {\bibinfo  {journal} {J. Phys. Soc. Jpn.}\ }\textbf {\bibinfo {volume}
  {84}},\ \bibinfo {pages} {023701} (\bibinfo {year} {2015})}\BibitemShut
  {NoStop}%
\bibitem [{\citenamefont {Takemura}\ \emph {et~al.}(2015)\citenamefont
  {Takemura}, \citenamefont {Takemori},\ and\ \citenamefont
  {Koga}}]{takemura15}%
  \BibitemOpen
  \bibfield  {author} {\bibinfo {author} {\bibfnamefont {S.}~\bibnamefont
  {Takemura}}, \bibinfo {author} {\bibfnamefont {N.}~\bibnamefont {Takemori}},
  \ and\ \bibinfo {author} {\bibfnamefont {A.}~\bibnamefont {Koga}},\ }\href
  {\doibase 10.1103/PhysRevB.91.165114} {\bibfield  {journal} {\bibinfo
  {journal} {Phys. Rev. B}\ }\textbf {\bibinfo {volume} {91}},\ \bibinfo
  {pages} {165114} (\bibinfo {year} {2015})}\BibitemShut {NoStop}%
\bibitem [{\citenamefont {Watanabe}\ and\ \citenamefont
  {Miyake}(2016)}]{watanabe16}%
  \BibitemOpen
  \bibfield  {author} {\bibinfo {author} {\bibfnamefont {S.}~\bibnamefont
  {Watanabe}}\ and\ \bibinfo {author} {\bibfnamefont {K.}~\bibnamefont
  {Miyake}},\ }\href {\doibase 10.7566/JPSJ.85.063703} {\bibfield  {journal}
  {\bibinfo  {journal} {J. Phys. Soc. Jpn.}\ }\textbf {\bibinfo {volume}
  {85}},\ \bibinfo {pages} {063703} (\bibinfo {year} {2016})}\BibitemShut
  {NoStop}%
\bibitem [{\citenamefont {Otsuki}\ and\ \citenamefont
  {Kusunose}(2016)}]{otsuki16}%
  \BibitemOpen
  \bibfield  {author} {\bibinfo {author} {\bibfnamefont {J.}~\bibnamefont
  {Otsuki}}\ and\ \bibinfo {author} {\bibfnamefont {H.}~\bibnamefont
  {Kusunose}},\ }\href {\doibase 10.7566/JPSJ.85.073712} {\bibfield  {journal}
  {\bibinfo  {journal} {J. Phys. Soc. Jpn.}\ }\textbf {\bibinfo {volume}
  {85}},\ \bibinfo {pages} {073712} (\bibinfo {year} {2016})}\BibitemShut
  {NoStop}%
\bibitem [{\citenamefont {{Shinzaki}}\ \emph {et~al.}(2016)\citenamefont
  {{Shinzaki}}, \citenamefont {{Nasu}},\ and\ \citenamefont
  {{Koga}}}]{shinzaki16}%
  \BibitemOpen
  \bibfield  {author} {\bibinfo {author} {\bibfnamefont {R.}~\bibnamefont
  {{Shinzaki}}}, \bibinfo {author} {\bibfnamefont {J.}~\bibnamefont {{Nasu}}},
  \ and\ \bibinfo {author} {\bibfnamefont {A.}~\bibnamefont {{Koga}}},\
  }\href@noop {} {\bibfield  {journal} {\bibinfo  {journal} {ArXiv e-prints}\ }
  (\bibinfo {year} {2016})},\ \Eprint {http://arxiv.org/abs/1606.06503}
  {arXiv:1606.06503 [cond-mat.str-el]} \BibitemShut {NoStop}%
\bibitem [{\citenamefont {Deguchi}\ \emph {et~al.}(2015)\citenamefont
  {Deguchi}, \citenamefont {Nakayama}, \citenamefont {Matsukawa}, \citenamefont
  {Imura}, \citenamefont {Tanaka}, \citenamefont {Ishimasa},\ and\
  \citenamefont {Sato}}]{deguchi15}%
  \BibitemOpen
  \bibfield  {author} {\bibinfo {author} {\bibfnamefont {K.}~\bibnamefont
  {Deguchi}}, \bibinfo {author} {\bibfnamefont {M.}~\bibnamefont {Nakayama}},
  \bibinfo {author} {\bibfnamefont {S.}~\bibnamefont {Matsukawa}}, \bibinfo
  {author} {\bibfnamefont {K.}~\bibnamefont {Imura}}, \bibinfo {author}
  {\bibfnamefont {K.}~\bibnamefont {Tanaka}}, \bibinfo {author} {\bibfnamefont
  {T.}~\bibnamefont {Ishimasa}}, \ and\ \bibinfo {author} {\bibfnamefont
  {N.~K.}\ \bibnamefont {Sato}},\ }\href {\doibase 10.7566/JPSJ.84.023705}
  {\bibfield  {journal} {\bibinfo  {journal} {J. Phys. Soc. Jpn.}\ }\textbf
  {\bibinfo {volume} {84}},\ \bibinfo {pages} {023705} (\bibinfo {year}
  {2015})}\BibitemShut {NoStop}%
\bibitem [{\citenamefont {Wong}\ \emph {et~al.}(1987)\citenamefont {Wong},
  \citenamefont {Lopdrup}, \citenamefont {Wagner}, \citenamefont {Shen},\ and\
  \citenamefont {Poon}}]{wong87}%
  \BibitemOpen
  \bibfield  {author} {\bibinfo {author} {\bibfnamefont {K.~M.}\ \bibnamefont
  {Wong}}, \bibinfo {author} {\bibfnamefont {E.}~\bibnamefont {Lopdrup}},
  \bibinfo {author} {\bibfnamefont {J.~L.}\ \bibnamefont {Wagner}}, \bibinfo
  {author} {\bibfnamefont {Y.}~\bibnamefont {Shen}}, \ and\ \bibinfo {author}
  {\bibfnamefont {S.~J.}\ \bibnamefont {Poon}},\ }\href {\doibase
  10.1103/PhysRevB.35.2494} {\bibfield  {journal} {\bibinfo  {journal} {Phys.
  Rev. B}\ }\textbf {\bibinfo {volume} {35}},\ \bibinfo {pages} {2494}
  (\bibinfo {year} {1987})}\BibitemShut {NoStop}%
\bibitem [{\citenamefont {Wagner}\ \emph {et~al.}(1988)\citenamefont {Wagner},
  \citenamefont {Biggs}, \citenamefont {Wong},\ and\ \citenamefont
  {Poon}}]{wagner88}%
  \BibitemOpen
  \bibfield  {author} {\bibinfo {author} {\bibfnamefont {J.~L.}\ \bibnamefont
  {Wagner}}, \bibinfo {author} {\bibfnamefont {B.~D.}\ \bibnamefont {Biggs}},
  \bibinfo {author} {\bibfnamefont {K.~M.}\ \bibnamefont {Wong}}, \ and\
  \bibinfo {author} {\bibfnamefont {S.~J.}\ \bibnamefont {Poon}},\ }\href
  {\doibase 10.1103/PhysRevB.38.7436} {\bibfield  {journal} {\bibinfo
  {journal} {Phys. Rev. B}\ }\textbf {\bibinfo {volume} {38}},\ \bibinfo
  {pages} {7436} (\bibinfo {year} {1988})}\BibitemShut {NoStop}%
\bibitem [{\citenamefont {Guidoni}\ \emph {et~al.}(1997)\citenamefont
  {Guidoni}, \citenamefont {Trich\'e}, \citenamefont {Verkerk},\ and\
  \citenamefont {Grynberg}}]{guidoni97}%
  \BibitemOpen
  \bibfield  {author} {\bibinfo {author} {\bibfnamefont {L.}~\bibnamefont
  {Guidoni}}, \bibinfo {author} {\bibfnamefont {C.}~\bibnamefont {Trich\'e}},
  \bibinfo {author} {\bibfnamefont {P.}~\bibnamefont {Verkerk}}, \ and\
  \bibinfo {author} {\bibfnamefont {G.}~\bibnamefont {Grynberg}},\ }\href
  {\doibase 10.1103/PhysRevLett.79.3363} {\bibfield  {journal} {\bibinfo
  {journal} {Phys. Rev. Lett.}\ }\textbf {\bibinfo {volume} {79}},\ \bibinfo
  {pages} {3363} (\bibinfo {year} {1997})}\BibitemShut {NoStop}%
\bibitem [{\citenamefont {Guidoni}\ \emph {et~al.}(1999)\citenamefont
  {Guidoni}, \citenamefont {D\'epret}, \citenamefont {di~Stefano},\ and\
  \citenamefont {Verkerk}}]{guidoni99}%
  \BibitemOpen
  \bibfield  {author} {\bibinfo {author} {\bibfnamefont {L.}~\bibnamefont
  {Guidoni}}, \bibinfo {author} {\bibfnamefont {B.}~\bibnamefont {D\'epret}},
  \bibinfo {author} {\bibfnamefont {A.}~\bibnamefont {di~Stefano}}, \ and\
  \bibinfo {author} {\bibfnamefont {P.}~\bibnamefont {Verkerk}},\ }\href
  {\doibase 10.1103/PhysRevA.60.R4233} {\bibfield  {journal} {\bibinfo
  {journal} {Phys. Rev. A}\ }\textbf {\bibinfo {volume} {60}},\ \bibinfo
  {pages} {R4233} (\bibinfo {year} {1999})}\BibitemShut {NoStop}%
\bibitem [{\citenamefont {Sanchez-Palencia}\ and\ \citenamefont
  {Santos}(2005)}]{sanchez05}%
  \BibitemOpen
  \bibfield  {author} {\bibinfo {author} {\bibfnamefont {L.}~\bibnamefont
  {Sanchez-Palencia}}\ and\ \bibinfo {author} {\bibfnamefont {L.}~\bibnamefont
  {Santos}},\ }\href {\doibase 10.1103/PhysRevA.72.053607} {\bibfield
  {journal} {\bibinfo  {journal} {Phys. Rev. A}\ }\textbf {\bibinfo {volume}
  {72}},\ \bibinfo {pages} {053607} (\bibinfo {year} {2005})}\BibitemShut
  {NoStop}%
\bibitem [{\citenamefont {Anderson}(1959)}]{anderson59}%
  \BibitemOpen
  \bibfield  {author} {\bibinfo {author} {\bibfnamefont {P.}~\bibnamefont
  {Anderson}},\ }\href {\doibase
  http://dx.doi.org/10.1016/0022-3697(59)90036-8} {\bibfield  {journal}
  {\bibinfo  {journal} {Journal of Physics and Chemistry of Solids}\ }\textbf
  {\bibinfo {volume} {11}},\ \bibinfo {pages} {26 } (\bibinfo {year}
  {1959})}\BibitemShut {NoStop}%
\bibitem [{\citenamefont {Penrose}(1974)}]{penrose74}%
  \BibitemOpen
  \bibfield  {author} {\bibinfo {author} {\bibfnamefont {R.}~\bibnamefont
  {Penrose}},\ }\href@noop {} {\bibfield  {journal} {\bibinfo  {journal} {Bull.
  Inst. Math. Appl.}\ }\textbf {\bibinfo {volume} {10}},\ \bibinfo {pages}
  {266} (\bibinfo {year} {1974})}\BibitemShut {NoStop}%
\bibitem [{\citenamefont {Micnas}\ \emph {et~al.}(1990)\citenamefont {Micnas},
  \citenamefont {Ranninger},\ and\ \citenamefont {Robaszkiewicz}}]{micnas90}%
  \BibitemOpen
  \bibfield  {author} {\bibinfo {author} {\bibfnamefont {R.}~\bibnamefont
  {Micnas}}, \bibinfo {author} {\bibfnamefont {J.}~\bibnamefont {Ranninger}}, \
  and\ \bibinfo {author} {\bibfnamefont {S.}~\bibnamefont {Robaszkiewicz}},\
  }\href {\doibase 10.1103/RevModPhys.62.113} {\bibfield  {journal} {\bibinfo
  {journal} {Rev. Mod. Phys.}\ }\textbf {\bibinfo {volume} {62}},\ \bibinfo
  {pages} {113} (\bibinfo {year} {1990})}\BibitemShut {NoStop}%
\bibitem [{\citenamefont {Bardeen}\ \emph {et~al.}(1957)\citenamefont
  {Bardeen}, \citenamefont {Cooper},\ and\ \citenamefont
  {Schrieffer}}]{bardeen57}%
  \BibitemOpen
  \bibfield  {author} {\bibinfo {author} {\bibfnamefont {J.}~\bibnamefont
  {Bardeen}}, \bibinfo {author} {\bibfnamefont {L.~N.}\ \bibnamefont {Cooper}},
  \ and\ \bibinfo {author} {\bibfnamefont {J.~R.}\ \bibnamefont {Schrieffer}},\
  }\href {\doibase 10.1103/PhysRev.108.1175} {\bibfield  {journal} {\bibinfo
  {journal} {Phys. Rev.}\ }\textbf {\bibinfo {volume} {108}},\ \bibinfo {pages}
  {1175} (\bibinfo {year} {1957})}\BibitemShut {NoStop}%
\bibitem [{\citenamefont {Metzner}\ and\ \citenamefont
  {Vollhardt}(1989)}]{metzner89}%
  \BibitemOpen
  \bibfield  {author} {\bibinfo {author} {\bibfnamefont {W.}~\bibnamefont
  {Metzner}}\ and\ \bibinfo {author} {\bibfnamefont {D.}~\bibnamefont
  {Vollhardt}},\ }\href {\doibase 10.1103/PhysRevLett.62.324} {\bibfield
  {journal} {\bibinfo  {journal} {Phys. Rev. Lett.}\ }\textbf {\bibinfo
  {volume} {62}},\ \bibinfo {pages} {324} (\bibinfo {year} {1989})}\BibitemShut
  {NoStop}%
\bibitem [{\citenamefont {Georges}\ \emph {et~al.}(1996)\citenamefont
  {Georges}, \citenamefont {Kotliar}, \citenamefont {Krauth},\ and\
  \citenamefont {Rozenberg}}]{georges96}%
  \BibitemOpen
  \bibfield  {author} {\bibinfo {author} {\bibfnamefont {A.}~\bibnamefont
  {Georges}}, \bibinfo {author} {\bibfnamefont {G.}~\bibnamefont {Kotliar}},
  \bibinfo {author} {\bibfnamefont {W.}~\bibnamefont {Krauth}}, \ and\ \bibinfo
  {author} {\bibfnamefont {M.~J.}\ \bibnamefont {Rozenberg}},\ }\href {\doibase
  10.1103/RevModPhys.68.13} {\bibfield  {journal} {\bibinfo  {journal} {Rev.
  Mod. Phys.}\ }\textbf {\bibinfo {volume} {68}},\ \bibinfo {pages} {13}
  (\bibinfo {year} {1996})}\BibitemShut {NoStop}%
\bibitem [{\citenamefont {Potthoff}\ and\ \citenamefont
  {Nolting}(1999)}]{potthoff99}%
  \BibitemOpen
  \bibfield  {author} {\bibinfo {author} {\bibfnamefont {M.}~\bibnamefont
  {Potthoff}}\ and\ \bibinfo {author} {\bibfnamefont {W.}~\bibnamefont
  {Nolting}},\ }\href {\doibase 10.1103/PhysRevB.59.2549} {\bibfield  {journal}
  {\bibinfo  {journal} {Phys. Rev. B}\ }\textbf {\bibinfo {volume} {59}},\
  \bibinfo {pages} {2549} (\bibinfo {year} {1999})}\BibitemShut {NoStop}%
\bibitem [{\citenamefont {Levine}\ and\ \citenamefont
  {Steinhardt}(1984)}]{levine84}%
  \BibitemOpen
  \bibfield  {author} {\bibinfo {author} {\bibfnamefont {D.}~\bibnamefont
  {Levine}}\ and\ \bibinfo {author} {\bibfnamefont {P.~J.}\ \bibnamefont
  {Steinhardt}},\ }\href {\doibase 10.1103/PhysRevLett.53.2477} {\bibfield
  {journal} {\bibinfo  {journal} {Phys. Rev. Lett.}\ }\textbf {\bibinfo
  {volume} {53}},\ \bibinfo {pages} {2477} (\bibinfo {year}
  {1984})}\BibitemShut {NoStop}%
\bibitem [{\citenamefont {Capone}\ \emph {et~al.}(2007)\citenamefont {Capone},
  \citenamefont {de' Medici},\ and\ \citenamefont {Georges}}]{capone07}%
  \BibitemOpen
  \bibfield  {author} {\bibinfo {author} {\bibfnamefont {M.}~\bibnamefont
  {Capone}}, \bibinfo {author} {\bibfnamefont {L.}~\bibnamefont {de' Medici}},
  \ and\ \bibinfo {author} {\bibfnamefont {A.}~\bibnamefont {Georges}},\ }\href
  {\doibase 10.1103/PhysRevB.76.245116} {\bibfield  {journal} {\bibinfo
  {journal} {Phys. Rev. B}\ }\textbf {\bibinfo {volume} {76}},\ \bibinfo
  {pages} {245116} (\bibinfo {year} {2007})}\BibitemShut {NoStop}%
\bibitem [{\citenamefont {Liebsch}\ and\ \citenamefont
  {Ishida}(2012)}]{liebsch12}%
  \BibitemOpen
  \bibfield  {author} {\bibinfo {author} {\bibfnamefont {A.}~\bibnamefont
  {Liebsch}}\ and\ \bibinfo {author} {\bibfnamefont {H.}~\bibnamefont
  {Ishida}},\ }\href {http://stacks.iop.org/0953-8984/24/i=5/a=053201}
  {\bibfield  {journal} {\bibinfo  {journal} {J. Phys.: Condens. Matter}\
  }\textbf {\bibinfo {volume} {24}},\ \bibinfo {pages} {053201} (\bibinfo
  {year} {2012})}\BibitemShut {NoStop}%
\bibitem [{\citenamefont {Sakai}\ \emph {et~al.}(2015)\citenamefont {Sakai},
  \citenamefont {Civelli}, \citenamefont {Nomura},\ and\ \citenamefont
  {Imada}}]{sakai15}%
  \BibitemOpen
  \bibfield  {author} {\bibinfo {author} {\bibfnamefont {S.}~\bibnamefont
  {Sakai}}, \bibinfo {author} {\bibfnamefont {M.}~\bibnamefont {Civelli}},
  \bibinfo {author} {\bibfnamefont {Y.}~\bibnamefont {Nomura}}, \ and\ \bibinfo
  {author} {\bibfnamefont {M.}~\bibnamefont {Imada}},\ }\href@noop {}
  {\bibfield  {journal} {\bibinfo  {journal} {Phys. Rev. B}\ }\textbf {\bibinfo
  {volume} {92}},\ \bibinfo {pages} {180503} (\bibinfo {year}
  {2015})}\BibitemShut {NoStop}%
\bibitem [{\citenamefont {Sakai}\ \emph {et~al.}(2016)\citenamefont {Sakai},
  \citenamefont {Civelli},\ and\ \citenamefont {Imada}}]{sakai16}%
  \BibitemOpen
  \bibfield  {author} {\bibinfo {author} {\bibfnamefont {S.}~\bibnamefont
  {Sakai}}, \bibinfo {author} {\bibfnamefont {M.}~\bibnamefont {Civelli}}, \
  and\ \bibinfo {author} {\bibfnamefont {M.}~\bibnamefont {Imada}},\ }\href
  {\doibase 10.1103/PhysRevLett.116.057003} {\bibfield  {journal} {\bibinfo
  {journal} {Phys. Rev. Lett.}\ }\textbf {\bibinfo {volume} {116}},\ \bibinfo
  {pages} {057003} (\bibinfo {year} {2016})}\BibitemShut {NoStop}%
\bibitem [{\citenamefont {Garg}\ \emph {et~al.}(2005)\citenamefont {Garg},
  \citenamefont {Krishnamurthy},\ and\ \citenamefont {Randeria}}]{garg05}%
  \BibitemOpen
  \bibfield  {author} {\bibinfo {author} {\bibfnamefont {A.}~\bibnamefont
  {Garg}}, \bibinfo {author} {\bibfnamefont {H.~R.}\ \bibnamefont
  {Krishnamurthy}}, \ and\ \bibinfo {author} {\bibfnamefont {M.}~\bibnamefont
  {Randeria}},\ }\href {\doibase 10.1103/PhysRevB.72.024517} {\bibfield
  {journal} {\bibinfo  {journal} {Phys. Rev. B}\ }\textbf {\bibinfo {volume}
  {72}},\ \bibinfo {pages} {024517} (\bibinfo {year} {2005})}\BibitemShut
  {NoStop}%
\bibitem [{\citenamefont {Toschi}\ \emph {et~al.}(2005)\citenamefont {Toschi},
  \citenamefont {Capone},\ and\ \citenamefont {Castellani}}]{toschi05}%
  \BibitemOpen
  \bibfield  {author} {\bibinfo {author} {\bibfnamefont {A.}~\bibnamefont
  {Toschi}}, \bibinfo {author} {\bibfnamefont {M.}~\bibnamefont {Capone}}, \
  and\ \bibinfo {author} {\bibfnamefont {C.}~\bibnamefont {Castellani}},\
  }\href {\doibase 10.1103/PhysRevB.72.235118} {\bibfield  {journal} {\bibinfo
  {journal} {Phys. Rev. B}\ }\textbf {\bibinfo {volume} {72}},\ \bibinfo
  {pages} {235118} (\bibinfo {year} {2005})}\BibitemShut {NoStop}%
\bibitem [{\citenamefont {Bauer}\ and\ \citenamefont {Hewson}(2009)}]{bauer09}%
  \BibitemOpen
  \bibfield  {author} {\bibinfo {author} {\bibfnamefont {J.}~\bibnamefont
  {Bauer}}\ and\ \bibinfo {author} {\bibfnamefont {A.~C.}\ \bibnamefont
  {Hewson}},\ }\href {http://stacks.iop.org/0295-5075/85/i=2/a=27001}
  {\bibfield  {journal} {\bibinfo  {journal} {EPL (Europhysics Letters)}\
  }\textbf {\bibinfo {volume} {85}},\ \bibinfo {pages} {27001} (\bibinfo {year}
  {2009})}\BibitemShut {NoStop}%
\bibitem [{\citenamefont {Koga}\ and\ \citenamefont {Werner}(2011)}]{koga11}%
  \BibitemOpen
  \bibfield  {author} {\bibinfo {author} {\bibfnamefont {A.}~\bibnamefont
  {Koga}}\ and\ \bibinfo {author} {\bibfnamefont {P.}~\bibnamefont {Werner}},\
  }\href {\doibase 10.1103/PhysRevA.84.023638} {\bibfield  {journal} {\bibinfo
  {journal} {Phys. Rev. A}\ }\textbf {\bibinfo {volume} {84}},\ \bibinfo
  {pages} {023638} (\bibinfo {year} {2011})}\BibitemShut {NoStop}%
\bibitem [{\citenamefont {Peters}\ and\ \citenamefont
  {Bauer}(2015)}]{peters15}%
  \BibitemOpen
  \bibfield  {author} {\bibinfo {author} {\bibfnamefont {R.}~\bibnamefont
  {Peters}}\ and\ \bibinfo {author} {\bibfnamefont {J.}~\bibnamefont {Bauer}},\
  }\href {\doibase 10.1103/PhysRevB.92.014511} {\bibfield  {journal} {\bibinfo
  {journal} {Phys. Rev. B}\ }\textbf {\bibinfo {volume} {92}},\ \bibinfo
  {pages} {014511} (\bibinfo {year} {2015})}\BibitemShut {NoStop}%
\bibitem [{\citenamefont {Ghosal}\ \emph {et~al.}(1998)\citenamefont {Ghosal},
  \citenamefont {Randeria},\ and\ \citenamefont {Trivedi}}]{ghosal98}%
  \BibitemOpen
  \bibfield  {author} {\bibinfo {author} {\bibfnamefont {A.}~\bibnamefont
  {Ghosal}}, \bibinfo {author} {\bibfnamefont {M.}~\bibnamefont {Randeria}}, \
  and\ \bibinfo {author} {\bibfnamefont {N.}~\bibnamefont {Trivedi}},\ }\href
  {\doibase 10.1103/PhysRevLett.81.3940} {\bibfield  {journal} {\bibinfo
  {journal} {Phys. Rev. Lett.}\ }\textbf {\bibinfo {volume} {81}},\ \bibinfo
  {pages} {3940} (\bibinfo {year} {1998})}\BibitemShut {NoStop}%
\bibitem [{\citenamefont {Ghosal}\ \emph {et~al.}(2001)\citenamefont {Ghosal},
  \citenamefont {Randeria},\ and\ \citenamefont {Trivedi}}]{ghosal01}%
  \BibitemOpen
  \bibfield  {author} {\bibinfo {author} {\bibfnamefont {A.}~\bibnamefont
  {Ghosal}}, \bibinfo {author} {\bibfnamefont {M.}~\bibnamefont {Randeria}}, \
  and\ \bibinfo {author} {\bibfnamefont {N.}~\bibnamefont {Trivedi}},\ }\href
  {\doibase 10.1103/PhysRevB.65.014501} {\bibfield  {journal} {\bibinfo
  {journal} {Phys. Rev. B}\ }\textbf {\bibinfo {volume} {65}},\ \bibinfo
  {pages} {014501} (\bibinfo {year} {2001})}\BibitemShut {NoStop}%
\bibitem [{\citenamefont {Jagannathan}\ \emph {et~al.}(2007)\citenamefont
  {Jagannathan}, \citenamefont {Szallas}, \citenamefont {Wessel},\ and\
  \citenamefont {Duneau}}]{jagannathan07}%
  \BibitemOpen
  \bibfield  {author} {\bibinfo {author} {\bibfnamefont {A.}~\bibnamefont
  {Jagannathan}}, \bibinfo {author} {\bibfnamefont {A.}~\bibnamefont
  {Szallas}}, \bibinfo {author} {\bibfnamefont {S.}~\bibnamefont {Wessel}}, \
  and\ \bibinfo {author} {\bibfnamefont {M.}~\bibnamefont {Duneau}},\ }\href
  {\doibase 10.1103/PhysRevB.75.212407} {\bibfield  {journal} {\bibinfo
  {journal} {Phys. Rev. B}\ }\textbf {\bibinfo {volume} {75}},\ \bibinfo
  {pages} {212407} (\bibinfo {year} {2007})}\BibitemShut {NoStop}%
\end{thebibliography}%

\end{document}